\documentclass[aps,prd,twocolumn,superscriptaddress,amsmath,amssymb,amsthm,nofootinbib,preprintnumbers]{revtex4}
\usepackage{amsmath}
\usepackage{graphicx}
\usepackage{epstopdf}
\usepackage{float}
\usepackage{hyperref}
\usepackage{subfigure}
\usepackage{color}
\usepackage[T1]{fontenc}
\usepackage[utf8]{inputenc}
\usepackage[toc,page]{appendix}
\usepackage[usenames,dvipsnames]{xcolor}
\usepackage[normalem]{ulem}
\usepackage{amssymb}
\usepackage{mathrsfs}
\usepackage{grffile}

\begin{document}

\title{Topology of black hole thermodynamics in Lovelock gravity}

\author{Ning-Chen Bai}
\email{bainingchen@stu.scu.edu.cn}
\affiliation{Center for Theoretical Physics, College of Physics, Sichuan University, Chengdu, 610065, China}

\author{Lei Li}
\email{lilei@swust.edu.cn}
\affiliation{Coherent Electronics Quantum
Optics Laboratory, Southwest University of Science and Technology, Mianyang, 621010, China}
 
\author{Jun Tao}
\email{taojun@scu.edu.cn}
\affiliation{Center for Theoretical Physics, College of Physics, Sichuan University, Chengdu, 610065, China}



\begin{abstract}
In this work, we present a convenient method to perform the topological analysis of black hole thermodynamics. Utilizing the spinodal curve, thermodynamic critical points of a black hole are endowed with a topological quantity, Brouwer degree, which reflects intrinsic properties of the system under smooth deformations. Specially, in our setup, it can be easily calculated without exact solution of critical points. This enables us to conveniently investigate the topological transition between different thermodynamic systems, and give a topological classification for them. In this framework, topology of Lovelock AdS black holes with spherical horizon geometry is explored. Results show that charged black holes in arbitrary dimensions can be classified into the same topology class, whereas the $d=7$ and $d \geq 8$ uncharged black holes are in different topology classes. Moreover, we revisit the relation between different phase structures of these black holes from the viewpoint of topology. Some general topological properties of critical points are also discussed.
\end{abstract}

\maketitle
\newpage 


\section{Introduction}\label{intro}
The study of black hole thermodynamics continues to be one of the most exciting
areas in gravitational theory. Of especial interest are the phase transitions in asymptotically anti–de Sitter (AdS) black holes, motivated by the straightforward definition of thermodynamic equilibrium and the possible interpretation in the context of gauge/gravity duality \cite{Maldacena:1997re,Gubser:1998bc,Witten:1998qj}. Early examples include the Hawking-Page phase transition occurring between thermal radiation and large AdS black hole \cite{Hawking:1982dh}, which can be interpreted as the confinement/deconfinement phase
transition of gauge field \cite{Witten:1998zw}, and the van der Waals like phase transition found between charged small and large AdS black holes \cite{Chamblin:1999tk,Chamblin:1999hg}. 

A topic of active research in recent years is the interpretation of the cosmological constant as the thermodynamic pressure \cite{Caldarelli:1999xj,Kastor:2009wy,Dolan:2010ha}. This perspective, known as the extend phase space, has shed new insights into the thermodynamics and phase transitions of AdS black holes, including understanding the Hawking–Page phase transition as a solid/liquid transition \cite{Kubiznak:2014zwa}, and strengthening the analogy between van der Waals fluids and charged AdS black holes \cite{Kubiznak:2012wp,Wei:2015iwa,Wei:2019uqg,Wei:2019yvs}. Plenty of novel phase behaviors are also discovered in this framework, such as the reentrant large/small/large black hole
phase transition \cite{Altamirano:2013ane,Dehyadegari:2017hvd}, triple points where small/large/intermediate black holes can coexist \cite{Altamirano:2013uqa, Wei:2014hba}, superfluid black holes \cite{Hennigar:2016xwd} admiting a $\lambda$-line phase transition reminiscent of the superfluid transition in liquid $^4$He, etc.

In spite of phase transitions differing in their forms, critical points always emerge and record crucial thermodynamic information. Locally, a second-order phase transition occurs at the critical point. Critical exponents can be derived from the behavior of physical quantities near this point, which are believed to be universal and related to general features of the physical system \cite{reichl1998modern}. For black hole systems, in most cases, they were found to have the standard set of critical exponents expected from mean field theory \cite{Kubiznak:2016qmn}. A special case is the isolated critical point from Lovelock gravity, which was shown to possess non-mean-field exponents \cite{Dolan:2014vba}. Globally, the presence of multiple critical points\footnote{Here we take the unphysical critical points (with negative pressure or temperature) into account.} generally implies intriguing phase behaviors, such as the reentrant phase transition (typically two critical points \cite{Altamirano:2013ane,Dehyadegari:2017hvd}), triple point (typically three critical points \cite{Altamirano:2013uqa, Wei:2014hba}), and in paticular the superfluid black hole which admits infinite critical points \cite{Hennigar:2016xwd}. Therefore, discovering the nature of critical points is quite valuable, which can provide additional insight into black hole thermodynamics and may also reveal more features about quantum gravity.

To this aim, topology has been introduced to the critical points of black hole thermodynamics \cite{Wei:2021vdx}. This approach begins by constructing a two-dimensional vector field with zero points designed to be critical points of the thermodynamic system. Following Duan’s $\phi$-mapping topological current theory \cite{Duan:2018rbd,Duan:1984ws}, one can assign a \textit{topological charge} for each critical point to reflect its local property. Concretely, the first-order phase transition can extend from the critical point with negative topological charge, but not from the one with positive topological charge. Moreover, topological charges at critical points can be added together to construct a \textit{total topological charge} for the thermodynamic system. This allows us to determine the global property of the thermodynamic system, and classify various thermodynamic systems into a few classes. Thermodynamics of several AdS black holes has been revisited and classified utilizing such an approach \cite{Wei:2021vdx,Yerra:2022alz,Yerra:2022eov}. It is also interesting to see that the isolated critical point can be interpreted as a topological phase transition of a ‘vortex/anti-vortex pair' \cite{Ahmed:2022kyv}.

In this paper, we present an alternative method to perform the topological analysis of thermodynamics and phase transitions of black holes. This method relies on a topological quantity \textit{Brouwer degree} \cite{dinca2021brouwer}, which is invariant under smooth deformations of the system and reflects system's intrinsic properties. Utilizing features of \textit{spinodal curve} \cite{Wei:2019yvs}, including its continuous differentiability and relation to critical points, we construct such a topological quantity for the thermodynamic system. Specially, as we will see, this quantity can be directly calculated by using a mathematical formula without exact solution of critical points. This enables us to conveniently probe the topological transition between different thermodynamic systems, and give them a topological classification. We shall employ this approach to explore the topological properties of AdS black holes in Lovelock gravity \cite{Lovelock:1971yv,Wang:2020eln}. On the one hand, these black holes possess rich phase behaviors \cite{Xu:2014tja,Frassino:2014pha,Hendi:2015soe}, providing an excellent arena for discovering topological features of black hole thermodynamics. On the other hand, the topology of black hole thermodynamics in  Gauss-Bonnet gravity, namely the 2nd-order Lovelock gravity, has been investigated \cite{Yerra:2022alz}. It would be interesting to examine whether the higher curvature gravity, 
such as the 3rd-order Lovelock gravity, would affect the topological properties of black hole thermodynamics.

The structure of the paper is as follows. In Sec. \ref{degree}, we give a brief introduction to the Brouwer degree. Then, by use of the spinodal curve, we relate the Brouwer degree to balck hole thermodynamics. In Sec. \ref{Lovelock}, the topology of Lovelock black holes' thermodynamics is investigated. The charged case and uncharged case are discussed separately. In Sec. \ref{pseudo}, further discussions on the relation between topological degree (charge) and critical point are given. Finally, we summarize and discuss our results in Sec. \ref{conclu}.

\section{Brouwer degree and spinodal curve}\label{degree}	
Consider an open and bounded set $X \subset \mathbb{R}^n$ with (at least) a $C^1$-smooth map $f: X \rightarrow \mathbb{R}^n$. Let $y \in f \backslash f(\partial X)$ be a regular value of $f$, then the set $f^{-1} (y)=\left\{x_1,x_2,\cdots\right\}$ with $x_n \in X$ has a finite number of points, such that $f(x_n)=y$. Suppose the Jacobian $J(x_n)=\text{det} (\partial f/ \partial x_n) \neq 0$, one can define a topological quantity, called the Brouwer
degree of the map \cite{dinca2021brouwer}
\begin{equation}
\operatorname{deg}(f,X,y)=\sum_{x_n \in f^{-1}(y)} \operatorname{sgn} J(x_n), \label{eq:Brouwer_degree}
\end{equation}
where sgn denotes the sign function
\begin{equation}
 \operatorname{sgn} (x)= \begin{cases}-1, & x<0 \\ 0, & x=0 \\ 1, & x>0\end{cases}.   
\end{equation}
This quantity is a topological characteristic of the map itself, which does not depend on the choice of the regular value $y$ and remains constant under continuous deformations of the map. 

\begin{figure}
\centering
\includegraphics[height=5cm]{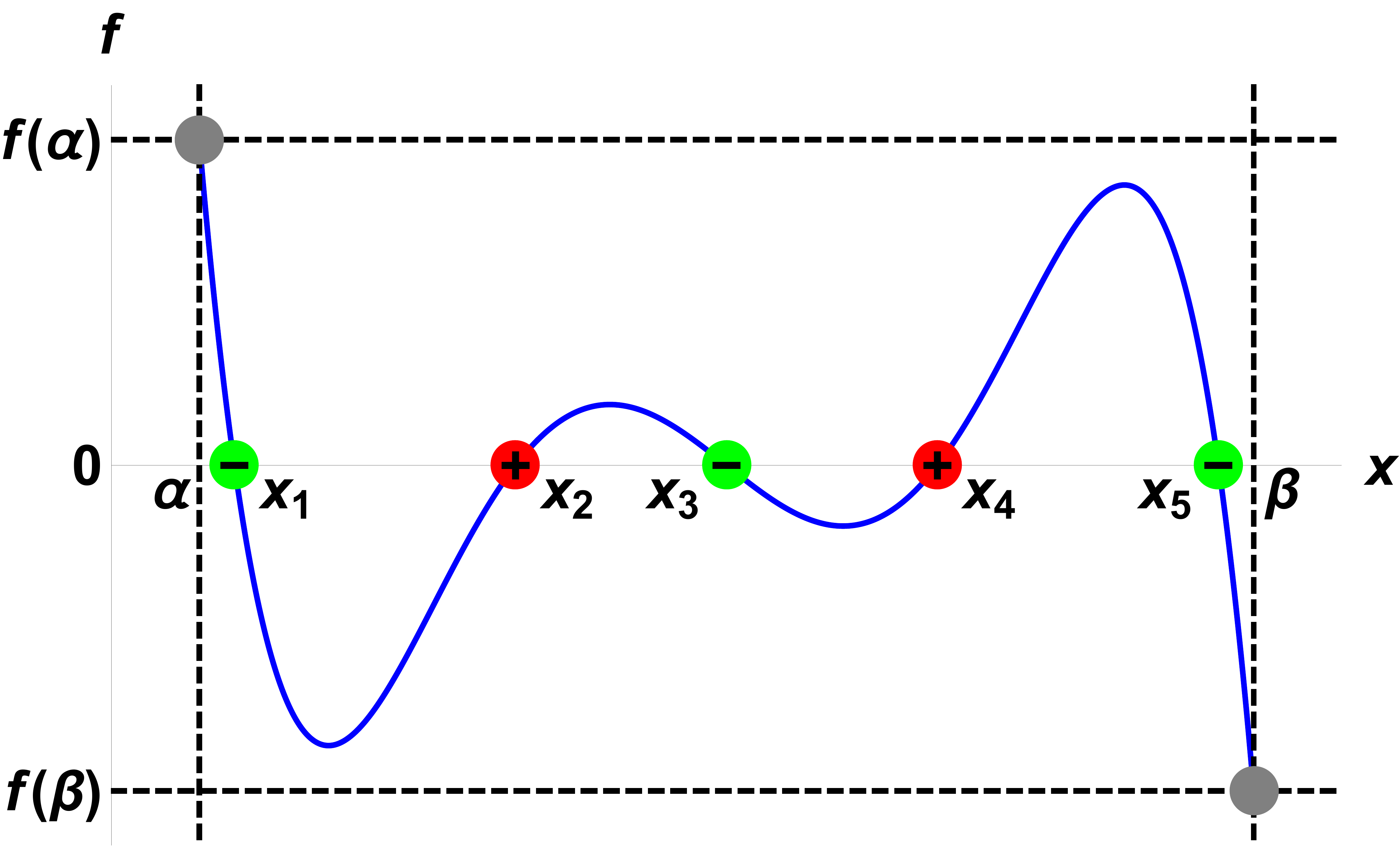}
\caption{\label{fig:one_dimension}A continuously differentiable function $f(x)$ with $x \in [\alpha,\beta]$, and $f(\alpha) \neq 0$, $f(\beta) \neq 0$.}
\end{figure}

Now we focus on the 1-dimensional case. Let $X=[\alpha,\beta]\subset \mathbb{R}$ and $f: X \rightarrow \mathbb{R}$ be a continuously differentiable function with $f(\alpha) \neq 0$ and $f(\beta) \neq 0$, and we choose $y=0$ such that $\{x_n\}=f^{-1} (y)$ be the set of zeroes of $f$, as shown in Fig. \ref{fig:one_dimension}. The non-zero Jacobian $J(x_n)$ now restricts $f'(x_n) \neq 0$, i.e., 0
is a regular value of $f$. Then the Brouwer degree (\ref{eq:Brouwer_degree}) can be expressed as
\begin{equation}
\operatorname{deg}(f,X,0)=\sum_{x_n \in f^{-1}(0)} \operatorname{sgn} f'(x_n). \label{eq:one_dimention_degree}
\end{equation}
Following Refs. \cite{Fu:2000pb,Cunha:2017qtt}, we can associate a topological charge $Q_{n} \equiv \operatorname{sgn} f'(x_n)$ for each zero point, and then sum over all contributions to construct the total topological charge (degree)
\begin{equation}
Q_{\text{total}}=\sum_{n} Q_{n}. \label{eq:one_dimention_degree}
\end{equation}
Note that the topological charge $Q_n$ can have two values, $-1$ or $+1$, according to the slope at the zero point $x_n$, as shown in Fig. \ref{fig:one_dimension}. The green and red points represent the zero points with negative and positive topological charges respectively. In the case of Fig. \ref{fig:one_dimension}, the total topological charge can be easily read as $Q_{\text{total}}=-1$.

Moreover, we notice that there is another simple way to calculate the total topological charge, which is accomplished by using the following formula \cite{liddell2015topological}:
\begin{equation}
\frac{1}{2} [\operatorname{sgn} f(\beta)-\operatorname{sgn} f(\alpha)]=\sum_{x_n \in f^{-1}(0)} \operatorname{sgn} f'(x_n). \label{eq:one_dimention_degree}
\end{equation}
From this equation, we can directly obtain the total topological charge by studying the asymptotic behavior of $f$, even if the zero points are undetermined. For the case of Fig. \ref{fig:one_dimension}, we have $Q_{\text{total}}=(-1-1)/2=-1$, which is consist with the result obtained before. Note that Eq. (\ref{eq:one_dimention_degree}) must be used with caution, which can only be applicable for a continuously differentiable function with non-zero boundary, see Ref. \cite{liddell2015topological} for more discussions.

To apply this tool to the topological study of thermodynamics, one has to endow the function $f$ and its zero points with specific physical significance. Recall that for a general state equation $T=T(V,P,z^i)$, the critical point can be identified with
\begin{equation}
\left(\frac{\partial T}{\partial V}\right)_{P, z^i}=0, \quad\left(\frac{\partial^{2} T}{\partial V^{2}}\right)_{P, z^i}=0.
\label{eq:T_critical_point}
\end{equation}
Using the first equation, one can eliminate the parameter $P$, and then get the spinodal curve $T_{\text{sp}}=T(V,z^i)$ \cite{Wei:2019yvs}. Now the condition (\ref{eq:T_critical_point}) turns into
\begin{equation}
\left(\partial_{V} T_{\text{sp}}\right)_{z^i}=0.
\label{eq:sp_derivative}
\end{equation}
Hence, we can let $f \equiv (\partial_{V} T_{\text{sp}})_{z^i}$, and thus zero points of $f$ exactly become critical points of the thermodynamic system. In this context, we can endow a topological charge for each critical point, and a total topological charge for the system to investigate the global properties.

\begin{figure}
\centering
\includegraphics[height=5cm]{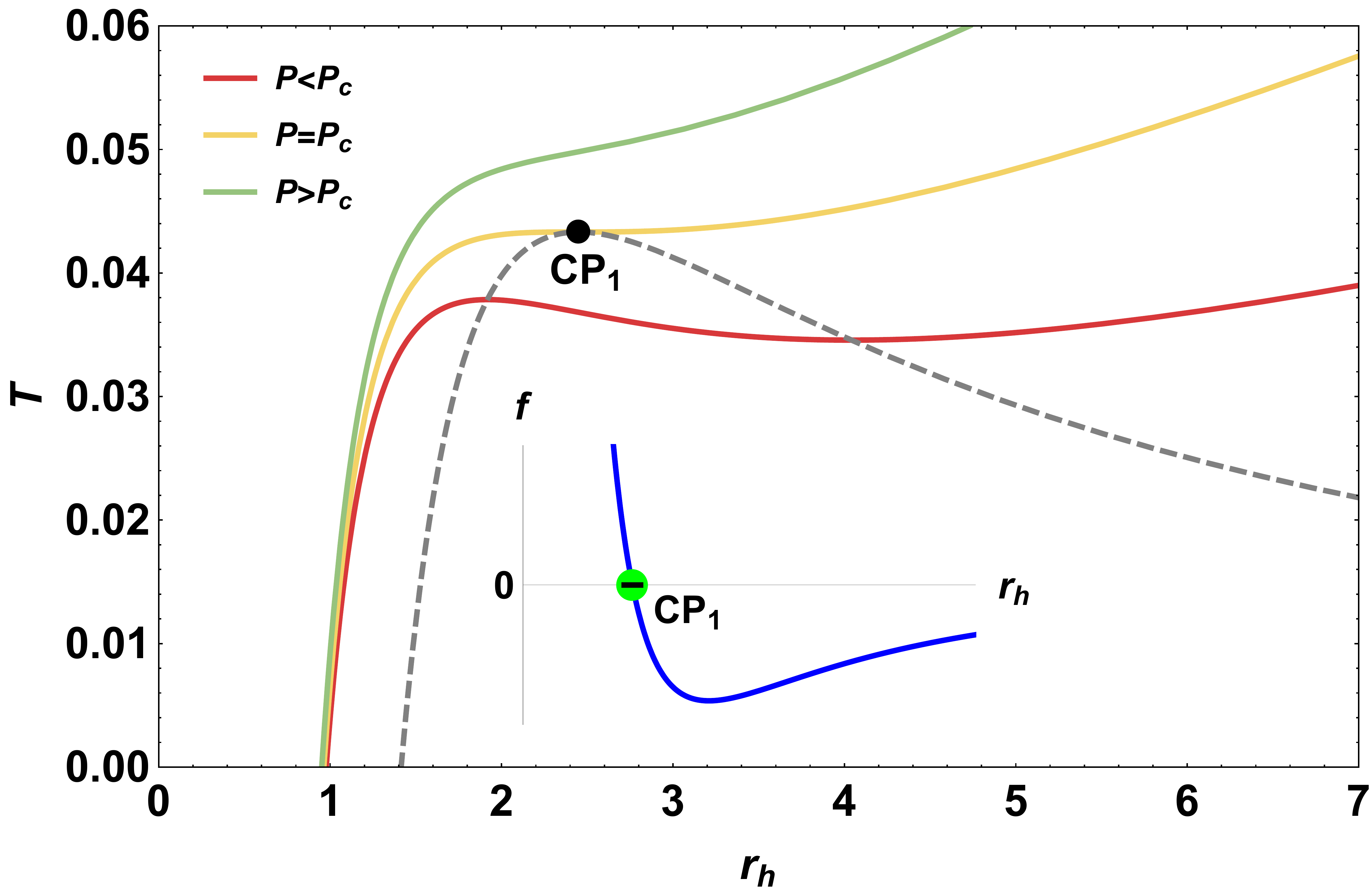}
\caption{\label{fig:charged_AdS}Isobaric curves and spinodal curve (gray dashed line) for the charged AdS black hole. Inset: $f \equiv (\partial_{r_{\text{h}}} T_{\text{sp}})_q$ vs $r_{\text{h}}$ diagram. We have set $d=4$ and $q=1$.}
\end{figure}

To make this claim more clearly, we first take a simple example---the charged AdS black hole system. Identifying the cosmological constant as the thermodynamic pressure $P$, the state equation in $d$-dimensional spacetime reads \cite{Gunasekaran:2012dq}
\begin{equation}
T=\frac{16 \pi r_{\mathrm{h}}^{2}\left(P-\frac{2 \pi q^{2} r_{\mathrm{h}}^{4-2 d}}{\omega_{d-2}^{2}}\right)+d (d-5)+6}{4 \pi(d-2) r_{\mathrm{h}}},
\end{equation}
where $r_{\text{h}}$ is the horizon radius, $q$ the electric charge, and $\omega_{d}=2 \pi^{(d+1) / 2} / \Gamma((d+1) / 2)$ the volume of a unit
$d$-sphere. After simple calculation, the spinodal curve can be obtained as
\begin{equation}
T_{\text{sp}}=\frac{1}{2 \pi  r_{\text{h}}} \left(-\frac{32 \pi ^2 q^2 r_{\text{h}}^{6-2 d}}{\omega _{d-2}^2}+d-3\right).
\end{equation}
Taking $d=4$, $q=1$ for example, we show the isobaric curves and the spinodal curve in Fig. \ref{fig:charged_AdS}. It is clear that the critical point $\text{CP}_1$ is exactly the extreme point of spinodal curve. Note that the condition for critical point $(\partial_{V} T_{\text{sp}})_{q}= 0$ is equal to $(\partial_{r_{\text{h}}} T_{\text{sp}})_q= 0$.  

Now we construct the function
\begin{equation}
f \equiv \left(\partial_{r_{\text{h}}} T_{\text{sp}}\right)_{q}=\frac{16 \pi  (2 d-5) q^2 r_{\text{h}}^{4-2 d}}{\omega _{d-2}^2}-\frac{d-3}{2 \pi  r_{\text{h}}^2},    
\end{equation}
and it is obvious that this function is continuously differentiable. It is also easy to see that for any $d \geq 4$ and $q >0$,
\begin{eqnarray}
 f(r_{\text{h}} \rightarrow 0^+) &\sim& \frac{(2 d-5) q^2 r_{\text{h}}^{4-2 d}}{\omega _{d-2}^2} \rightarrow +\infty,\\
 f(r_{\text{h}} \rightarrow +\infty)  &\sim& -\frac{d-3}{r_{\text{h}}^2} \rightarrow 0^-,
\end{eqnarray}
and thus $f$ admits non-zero boundary. For these features, we can directly use Eq. (\ref{eq:one_dimention_degree}) to calculate the total topological charge:
\begin{equation}
\begin{aligned}
Q_{\text{total}} &= \frac{1}{2} [\operatorname{sgn} f(r_{\text{h}} \rightarrow +\infty)-\operatorname{sgn} f(r_{\text{h}} \rightarrow 0^+)]\\
&= \frac{1}{2} (-1-1)=-1\,.
\end{aligned}
\end{equation}
Therefore we recover the $d=4$, $q=1$ result obtained in Ref. \cite{Wei:2021vdx} using a different method, and further show that for any $d \geq 4$ and $q >0$, the charged AdS black holes share the same topological charge $-1$. This also implies that there is \textit{at least} one critical point presented in the system, otherwise a zero topological charge will be obtained. Note that we capture this information without actually solving Eq. (\ref{eq:sp_derivative}).

To obtain the topological charge for each critical point, one can study the behavior of function $f$ at zero points. In the inset of Fig. \ref{fig:charged_AdS}, we display the $d=4$, $q=1$ case. Obviously, there is a zero point with negative slope, indicating a critical point $\text{CP}_{1}$ with the topological charge
\begin{equation}
Q_{\text{CP}_{1}} = \operatorname{sgn} f'(\text{CP}_{1}) =-1.
\end{equation}
In fact, using the condition (\ref{eq:T_critical_point}), one can show that charged AdS black holes only have one critical point, and thus $Q_{\text{CP}_{1}} \equiv Q_{\text{total}}=-1$. Moreover, since both $f=(\partial_{r_{\text{h}}} T_{\text{sp}})_q=0$ and $f'=(\partial_{r_{\text{h}},r_{\text{h}}} T_{\text{sp}})_q<0$ are satisfied, we conclude that a critical point with negative topological charge denotes a maximum point of spinodal curve. As shown in Fig. \ref{fig:charged_AdS}, near such a point, the stable black hole branches ($C_{P}=T\left(\partial_{S} T\right)^{-1}_{P,z^i}>0$) are on both sides, while the unstable black hole branch is in the middle, and one can draw a first-order phase transition line in $T-V$ plane among these stable black hole branches using Maxwell’s equal area law. However, whether such a phase transition can actually occur should be carefully examined by the free energy. This will be discussed further in Sec. \ref{pseudo}. 

On the contrary, a critical point with positive topological charge denotes a minimum point of spinodal curve. Near this point, the unstable black hole branches are on both sides, whereas the stable black hole branch is in the middle, and thus one can not draw a first-order phase transition line.

It is worth noting that in the above discussion, we suppose $f'(x_n) \neq 0$. In this case, it is obvious that the adjacent critical points must have opposite topological charges, as shown in Fig. \ref{fig:one_dimension}. Under the continuous deformations of $f$, if $f'(x_j) = 0$ is satisfied for some zero point $x_j$, there will be a critical point with zero slope. One can treat this point as the result of the \textit{annihilation} of adjacent critical points, and endow it with a zero topological charge. Since the total topological charge does not change in this process, Eq. (\ref{eq:one_dimention_degree}) is still applicable.

So far, we have constructed a convenient method to investigate the topology of black hole thermodynamics. The physical significance of the topological charge has also been clearly described in the context of charged AdS black holes. In the following sections, we shall focus on a more complex system---Lovelock black holes, in which more interesting topological properties will be disclosed.

\section{Thermodynamic topology of Lovelock black holes}\label{Lovelock}
The action of the Lovelock gravity in $d$-dimensional spacetime with a Maxwell field is given by \cite{Lovelock:1971yv}
\begin{equation}
I =  \frac{1}{16\pi G_N} \int d^dx  \sqrt{-g}\Bigl({\sum_{k=0}^{k_{max}} }\hat{\alpha}_{\left(k\right)}\mathcal{L}^{\left(k\right)} - 4\pi G_N F_{ab}F^{ab}\Bigr),
\label{eq:Loveaction}
\end{equation}
where $\hat{\alpha}_{\left(k\right)}$ are the Lovelock coupling constants and $\mathcal{L}^{\text{\ensuremath{\left(k\right)}}}$ are the $2k$-dimensional Euler densities, given by the contraction of $k$ powers of the Riemann tensor
\begin{equation}
\mathcal{L}^{\left(k\right)}=\frac{1}{2^{k}} \delta^{(k)} R_{a_{1}b_{1}}^{\quad c_{1}d_{1}}\ldots R_{a_{k}b_{k}}^{\quad c_{k}d_{k}},
\end{equation}
where $\delta^{(k)}=\delta_{c_{1}d_{1}\ldots c_{k}d_{k}}^{a_{1}b_{1}\ldots a_{k}b_{k}}$ is totally antisymmetric in both sets of indices. The $\mathcal{L}^{\left(0\right)}$, $\mathcal{L}^{\left(1\right)}$ and $\mathcal{L}^{\left(2\right)}$ correspond to the cosmological constant term, Einstein--Hilbert term and quadratic Gauss--Bonnet term, respectively. The integer $k_{max}=\left[\frac{d-1}{2}\right]$ restricts that only for $d>2k$, $\mathcal{L}^{\left(k\right)}$ contributes to the equations of motion.

Considering the charged AdS black holes in Lovelock gravity with static spherically symmetric, the ansatz is given by
\begin{eqnarray}
\label{solution}
ds^{2} &=& -g\left(r\right)dt^{2}+g\left(r\right)^{-1}dr^{2}+r^{2}d\Omega_{(\kappa) d-2}^{2},\\
F &=& \frac{Q}{r^{d-2}} dt\wedge dr, 
\end{eqnarray}
where $Q$ denotes the electric charge, $d\Omega_{(\kappa) d-2}^{2}$ is the line element of a $\left( d-2 \right)$-dimensional compact space with volume denoted by $\Sigma_{d-2}^{(\kappa)}$ and constant curvature $(d-2)(d-3)\kappa$, in which  $\kappa=0,+1,-1$ correspond to flat, spherical, hyperbolic black hole horizon geometries respectively.

For this ansatz, the field equations derived from Eq. (\ref{eq:Loveaction}) can be reduced to \cite{Boulware:1985wk,Wheeler:1985qd,Cai:2003kt}
\begin{eqnarray}
\sum_{k=0}^{k_{max}}\alpha_{k} \left(\frac{\kappa-g}{r^2}\right)^{k}&=&\frac{16\pi G_NM}{(d-2)\Sigma_{d-2}^{(\kappa)}r^{d-1}}\nonumber\\
&&-\frac{8\pi G_N Q^2}{(d-2)(d-3)}\frac{1}{r^{2d-4}}.
\end{eqnarray}
where
\begin{eqnarray}
\alpha_{0}&=&\frac{\hat{\alpha}_{(0)}}{\left(d-1\right)\left(d-2\right)}\,,\quad{\alpha}_{1}={\hat \alpha}_{(1)},\nonumber\\
\alpha_{k}&=&\hat \alpha_{(k)}\prod_{n=3}^{2k}\left(d-n\right)  {\quad\mbox{for}\quad  k\geq2},
\end{eqnarray}
and $M$ is the ADM mass of the black hole. To avoid the possible solutions with naked singularities \cite{Myers:1988ze} and conform to the string tension explanation in heterotic string theory, we focus on the $\alpha_k>0$ case in the followings. Specially, to recover general relativity in the small curvature limit, $\alpha_1$ is set to be $1$.

By using the Hamiltonian formalism, the thermodynamic quantities of the black hole, including the black hole mass $M$, the temperature $T$, the entropy $S$ and the electric potential $\Phi$, are calculated as \cite{Cai:2003kt}
\begin{widetext}
\begin{eqnarray}
M &=& \frac{\Sigma_{d-2}^{(\kappa)}\left(d-2\right)}{16\pi G_N}\sum_{k=0}^{k_{max}}\alpha_{k}\kappa^kr_+^{d-1-2k}+\frac{\Sigma_{d-2}^{(\kappa)}}{2(d-3)}\frac{Q^2}{r_+^{d-3}},\\
T &=& \frac{1}{4\pi r_+ D(r_+)}\Bigl[\sum_k\kappa\alpha_k(d-2k-1)\Bigl(\frac{\kappa}{r_+^2}\Bigr)^{k-1}-\frac{8\pi G_N Q^2}{(d-2)r_+^{2(d-3)}}\Bigr], \label{eq:T}\\
S &=& \frac{\Sigma_{d-2}^{(\kappa)}\left(d-2\right)}{4G_N}\sum_{k=0}^{k_{max}}\frac{k\kappa^{k-1}\alpha_{k}r_+^{d-2k}}{d-2k},\\
\Phi&=&\frac{\Sigma_{d-2}^{(\kappa)}Q}{(d-3)r_+^{d-3}}, \label{eq:Entropy}
\end{eqnarray}
\end{widetext}
where $r_{+}$ is the horizon radius determined as the
largest root of $f \left(r_{+}\right)=0$, and
\begin{equation}
{D(r_+)=\sum_{k=1}^{k_{max}}k\alpha_{k}\left(\kappa r_{+}^{-2}\right)^{k-1}.} 
\end{equation}

Identifying the cosmological constant $\Lambda=-\hat{\alpha}_{0}/2$ as the thermodynamic pressure,
\begin{equation}
P = -\frac{\Lambda}{8\pi G_N} = \frac{\hat{\alpha}_{0}}{16\pi G_N},
\end{equation}
the extended first law of black hole thermodynamics reads \cite{Kastor:2010gq,Frassino:2014pha}
\begin{eqnarray}
\delta M=T \delta S+V \delta P+\sum_{k=1}^{K} \Psi^{(k)} \delta \alpha_{k},
\end{eqnarray}
where $\Psi^{(k)}$ represents the quantity conjugate to $\alpha_{k}$,
\begin{equation}
\Psi^{(k)}=\frac{\Sigma_{d-2}^{(\kappa)}(d-2)}{16\pi G_N}\kappa^{k-1}{r^{d-2k}_+}\left[\frac{\kappa}{r}-\frac{4\pi kT}{d-2k}\right],
\end{equation}
and $V$ is the thermodynamic volume conjugate to $P$,
\begin{equation}
V=\frac{16\pi G_N \Psi^{(0)}}{(d-1)(d-2)}.
\end{equation}

Working in an ensemble that fixes $\alpha_k$ for $k\geq 1$, in terms of the following dimensionless quantities \cite{Frassino:2014pha}:
\begin{eqnarray}
&Q= \frac{q}{\sqrt{2}} \alpha_{3}^{\frac{d-3}{4}}, \quad {\alpha}=\frac{\alpha_2}{\sqrt{\alpha_3}}, \quad m=\frac{16\pi M}{(d-2)\Sigma_{d-2}^{(\kappa)}\alpha_3^{\frac{d-3}{4}}},& \nonumber\\
&r_+ = v\alpha_{3}^{\frac{1}{4}},\quad  T=\frac{t\alpha_{3}^{-\frac{1}{4}}}{d-2}, \quad p=4 \sqrt{\alpha_{3}} P,& 
\end{eqnarray}
one can reinterpret (\ref{eq:T}) as the state equation for 3rd-order Lovelock $U(1)$ charged black holes,
\begin{eqnarray}
t &=& \frac{1}{4 \pi  v \left(2 \alpha  \kappa  v^2+v^4+3\right)} \Bigl[-4 \pi  q^2 v^{10-2 d}\nonumber\\
&& + (d-7) (d-2) \kappa +\alpha  (d-5) (d-2) v^2 \nonumber\\
&&+(d-3) (d-2) \kappa  v^4+4 p \pi v^6\Bigr]\,, \label{eq:t_Lovelock}
\end{eqnarray}
and the condition (\ref{eq:T_critical_point}) for critical points now becomes
\begin{eqnarray}\label{cp}
\left(\frac{\partial t}{\partial v}\right)_{p,q,\alpha}=0,\quad \left(\frac{\partial^2 t}{\partial v^2}\right)_{p,q,\alpha}=0. \label{eq:critical_point_Lovelock}
\end{eqnarray}

The possible phase transitions can be investigated based on the behavior of the Gibbs free energy in the `canonical ensemble', given by \cite{Kastor:2010gq,Kastor:2011qp}
\begin{eqnarray}\label{G}
G=M-TS=G(P,T, Q,\alpha_1,\dots, \alpha_{k_{max}})\,.
\end{eqnarray}
with the dimensionless  counterpart \cite{Frassino:2014pha}
\begin{widetext}
\begin{eqnarray}\label{gLovelock}
g(t,p,q, \alpha)&=& \frac{1}{\Sigma_{d-2}^{(\kappa)}} \alpha_3^{\frac{3-d}{4}} G =-\frac{1}{16\pi(3+2\alpha\kappa v^2+v^4)}\Bigl[\frac{4\pi p v^{d+3}}{(d-1)(d-2)}
-\kappa v^{d+1}+\frac{24\pi\kappa p \alpha v^{d+1}}{(d-1)(d-4)}-\frac{\alpha v^{d-1}(d-8)}{d-4}\nonumber\\
&&+\frac{60\pi p v^{d-1}}{(d-1)(d-6)}-\frac{2\kappa \alpha^2 v^{d-3}(d-2)}{d-4}
+\frac{4\kappa v^{d-3}(d+3)}{d-6}-\frac{3\alpha v^{d-5}(d-2)(d-8)}{(d-4)(d-6)}-\frac{3\kappa v^{d-7}(d-2)}{d-6}\Bigr]\nonumber\\
&&+\frac{q^2}{4(3+2\alpha\kappa v^2+v^4)(d-3)v^{d-3}}
\Bigl[\frac{v^4(2d-5)}{d-2}+\frac{2\alpha\kappa(2d-7)v^2}{d-4}+\frac{3(2d-9)}{d-6}\Bigr].
\end{eqnarray}
\end{widetext}
The stable state corresponds to the global minimum of this quantity for its fixed parameters $t, p, q$ and $\alpha$.

For $\kappa=0$, one can find there is no critical point, which means that any planar black holes of higher-order Lovelock gravity in arbitrary dimensions do not exhibit critical behavior. Thus there is no urge to study its thermodynamic topology. On the other hand, the thermodynamic topology for the $\kappa=-1$ case has been investigated and intriguing result has been given; the isolated critical point---a peculiar thermodynamic critical point that occurs in the phase diagram of hyperbolic black holes, can be interpreted as a topological phase transition of a ‘vortex/anti-vortex
pair’ \cite{Ahmed:2022kyv}. In what follows, we shall concentrate on the thermodynamic topology of $\kappa=+1$ case, in which rich phase behaviors such as triple points and reentrant phase transitions exist, and thus interesting topological properties would be expected.

\subsection{Charged Case}
We first study the topology of thermodynamics in the charged case, i.e., $q > 0$. From Eq. (\ref{eq:t_Lovelock}) and the first equation of (\ref{eq:critical_point_Lovelock}), the spinodal curve is given by
\begin{eqnarray}
t_{\text{sp}} &=& \frac{d-2}{2 \pi  v^{2 d + 1} \left(6 \alpha  v^2+v^4+15\right)} \Bigl[-4 \pi  q^2 v^{10} + 3 ( d -7 ) v^{2 d} \nonumber\\
&&+ 2 ( d-5 ) \alpha v^{2 d+2}+(d -3)  v^{2 d+4}\Bigr]\,, \label{eq:tsp_charged_Lovelock}
\end{eqnarray}
and then we can define
\begin{widetext}
\begin{eqnarray}
f &\equiv& \left(\partial_{v} t_{\text{sp}}\right)_{q,\alpha} = -\frac{d-2}{2 \pi  v^{2 d+2} \left(6 \alpha  v^2+v^4+15\right)^2}\Bigl[60 \pi  (9-2 d) q^2 v^{10} + 24 \pi  \alpha  (7-2 d) q^2 v^{12}+4 \pi  (5-2 d) q^2 v^{14} \nonumber\\
&&+45 (d-7) v^{2 d} +4 \alpha  (6 d-57) v^{2 d+2} +6 \left(-10 \alpha ^2+2 \alpha ^2 d-5 d+5\right) v^{2 d+4} -12 \alpha  v^{2 d+6} +(d-3) v^{2 d+8}\Bigr]\,. \label{eq:f_charged_Lovelock}
\end{eqnarray}
\end{widetext}
Obviously, this function is continuously differentiable. In addition, for any $d \geq 7$ and $q > 0$, we have
\begin{eqnarray}
f(v \rightarrow 0^+) &\sim& \frac{(d-2)  (2 d-9) q^2 }{  v^{2 d-8} \left(6 \alpha  v^2+v^4+15\right)^2} \rightarrow +\infty, \nonumber\\
f(v \rightarrow +\infty)  &\sim& -\frac{ (d-2)  (d-3) v^{6}}{\left(6 \alpha  v^2+v^4+15\right)^2} \rightarrow 0^-,
\end{eqnarray}
hence $f$ admits non-zero boundary. By use of Eq. (\ref{eq:one_dimention_degree}), the total topological charge can be directly calculated as
\begin{eqnarray}
Q_{\text{total}} &=& \frac{1}{2} [\operatorname{sgn} f(v \rightarrow +\infty)-\operatorname{sgn} f(v \rightarrow 0^+)]\nonumber\\
&=& \frac{1}{2} (-1-1)=-1\,.
\end{eqnarray}
Since this result does not depend on the values of parameters $(q, \alpha)$ and dimension $d$, we conclude that charged Lovelock AdS black holes with spherical horizon geometry share the same topological charge, indicating they can be classified into the same topology class. More interestingly, they belong to the same topology class with the charged AdS black holes, which may imply some similarities between their thermodynamics. 

\begin{figure}
\centering
\includegraphics[height=5cm]{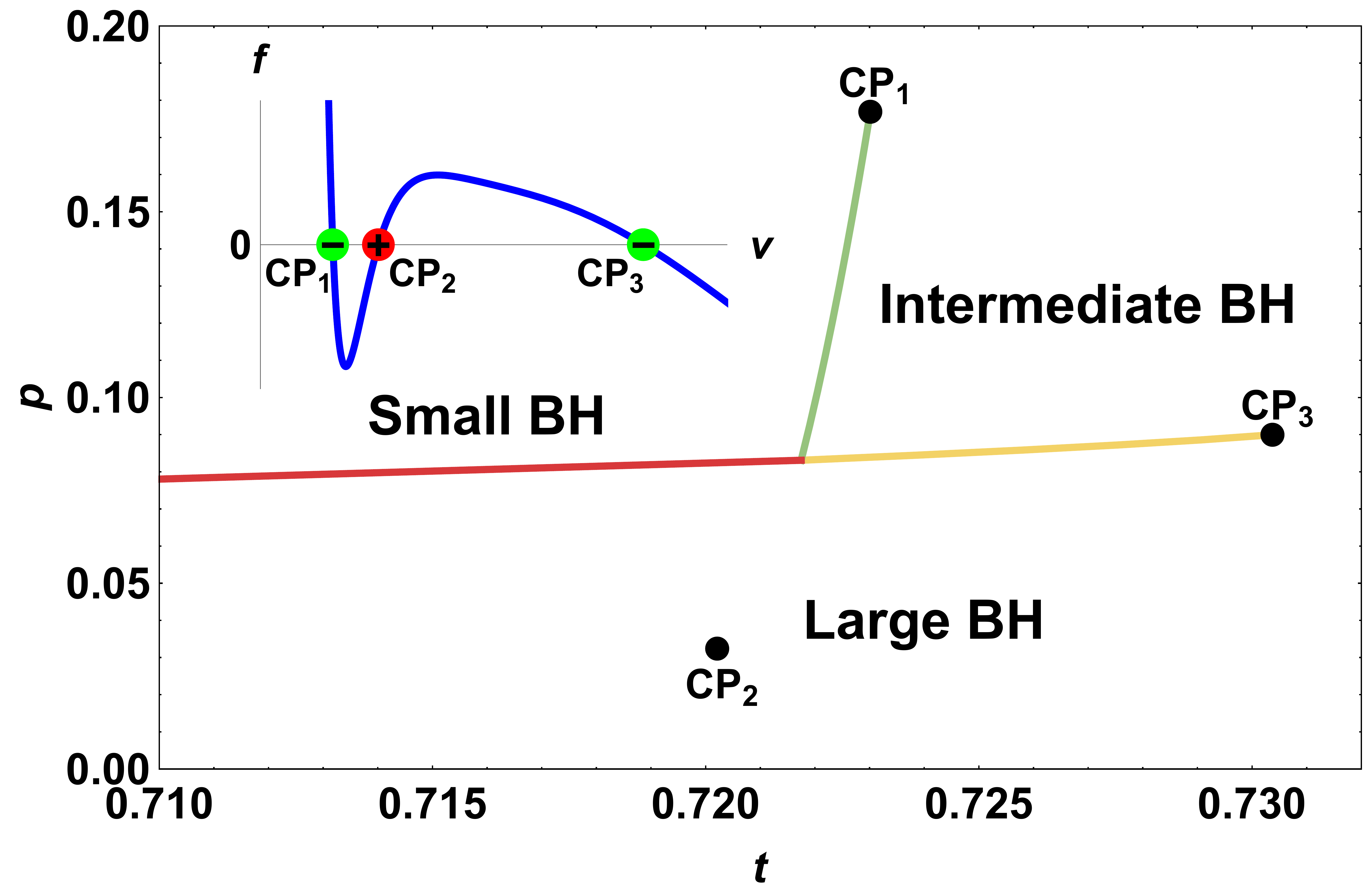}
\caption{\label{fig:P_T_charged} Small/intermediate/large black phase structure for the charged AdS Lovelock black hole with spherical horizon geometry. Inset: $f \equiv (\partial_{v} T_{\text{sp}})_{q,\alpha}$ vs $v$ diagram. We have set $(d,\alpha,q)=(8,2.8,0.0175)$.}
\end{figure}

Actually, in the $d=7$ case, one can find that the equation of state only admits one critical point, and the system exhibits the typical small/large black hole phase structure. While in the $d=8$ case, the state equation displays one or three critical points (including the unphysical ones) in appropriate parameter ranges, and a triple point may arise when a small charge $q$ is added to the black hole, at which the small, large and intermediate black holes can coexist together. Taking $\alpha=2.8$ and $q=0.0175$ for example, we display such a phase structure in Fig. \ref{fig:P_T_charged}. The invariant total topological charge shown above thus suggests that the small/intermediate/large black hole phase structure can be interpreted as the \textit{topological transformation} of small/large black hole phase structure. According to the behavior of $f$ shown in the inset of Fig. \ref{fig:P_T_charged}, it is easy to verify that the total topological charge indeed takes $-1$:
\begin{eqnarray}
Q_{\text{total}} &=& Q_{\text{CP}_1}+Q_{\text{CP}_2}+Q_{\text{CP}_3}\nonumber\\
&=&-1+1-1=-1.   
\end{eqnarray}

\begin{figure}
\centering
\includegraphics[height=5cm]{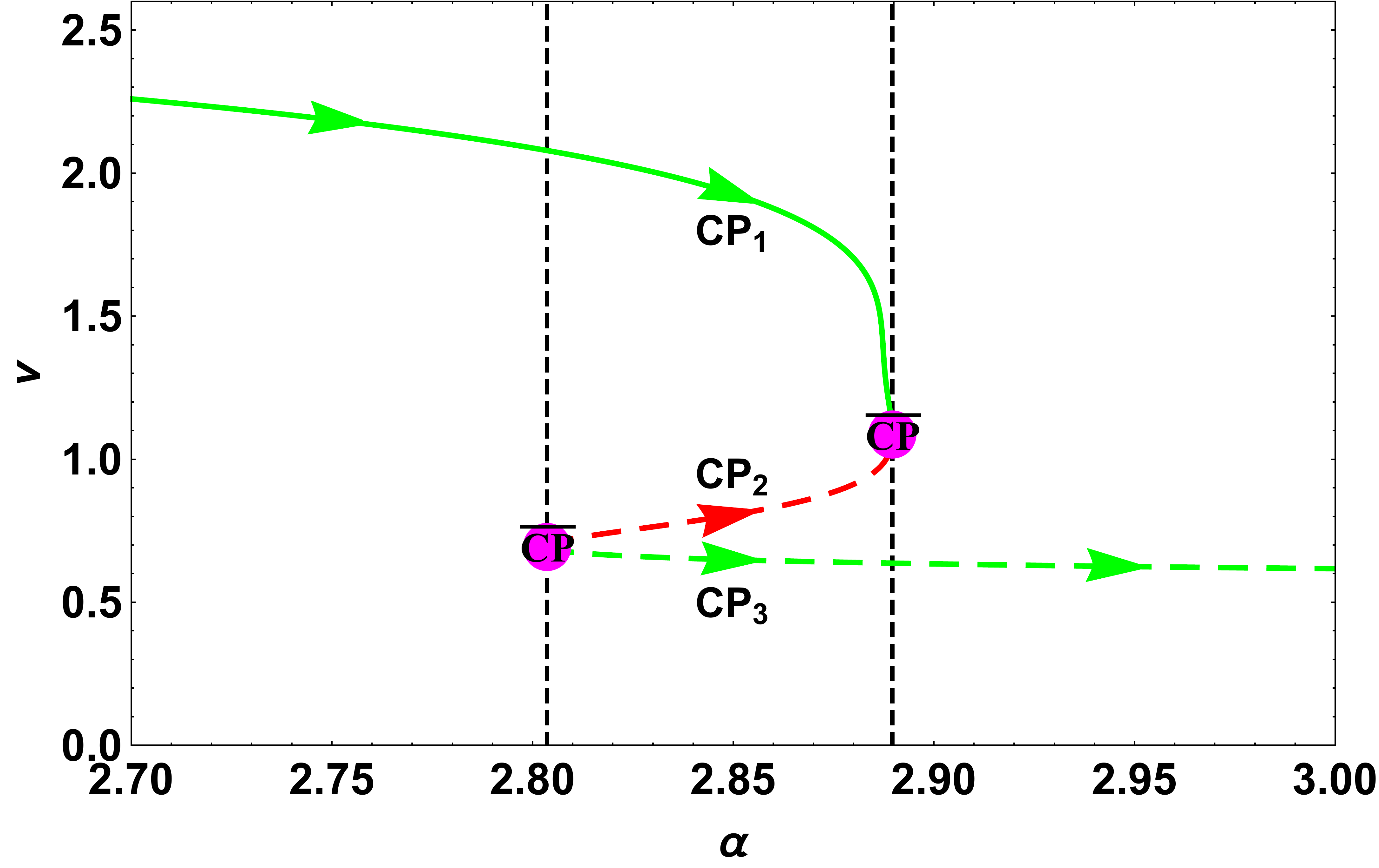}
\caption{\label{fig:Charged_Creation} Pair creation and annihilation of critical points for the charged AdS Lovelcock black hole. The green and red curves represent the branches with negative and positive topological charges, respectively. The black dashed line on the left denotes $\alpha_1 \approx 2.804$, while the right one denotes $\alpha_2 \approx 2.890$. The arrows refer to the direction of increasing $\alpha$. We have set $d=8$ and $q=0.022$.}
\end{figure}

Meanwhile, the invariant total topological charge also indicates that critical points must \textit{emerge}, \textit{or annihilate in pairs}, between the ones with \textit{opposite} topological charges. Similar phenomenons can also be seen in Refs. \cite{Fu:2000pb,Cunha:2017qtt,Duan:2003aj,Cao:2008zzc,Wei:2022mzv,Wei:2022dzw,Yerra:2022coh}. To actually observe such a behavior, we take $d=8$ and consider a process of varying $\alpha$ and fixing $q=0.022$. In this process, $\alpha$ can be treated as a `time evolution factor' of the system. The critical points are numerically solved using condition (\ref{eq:critical_point_Lovelock}), and the corresponding topological charge for each critical point is obtained according to the slope of $f$. Results are summarized in Fig. \ref{fig:Charged_Creation}.  When $\alpha < \alpha_1 \approx 2.804$, there is only one critical point $\text{CP}_1$ with topological charge $-1$. When $\alpha$ goes exactly to $\alpha_1$, besides the original $\text{CP}_1$, we observe a critical point $\overline{\mathrm{CP}}$ with zero slope of $f$, i.e., zero topological charge. Further increasing $\alpha$, $\text{CP}_1$ still exists, while $\overline{\mathrm{CP}}$ generates two new critical points:
\begin{equation}
\overline{\mathrm{CP}} \rightarrow \text{CP}_2 + \text{CP}_3,    
\end{equation}
in which $\text{CP}_2$ has topological charge $+1$, whereas $\text{CP}_3$ has topological charge $-1$. With the increasing of $\alpha$, these two critical points move away from each other, while $\text{CP}_1$ and $\text{CP}_2$ get closer. When $\alpha$ goes to $\alpha_2 \approx 2.890$, we observe a reverse process---two critical points merge into one critical point:
\begin{equation}
\text{CP}_1 + \text{CP}_2 \rightarrow \overline{\mathrm{CP}},    
\end{equation}
where again $\overline{\mathrm{CP}}$ has zero topological charge. Beyond $\alpha_2$, $\overline{\mathrm{CP}}$ also disappears, and only $\text{CP}_3$ is left. Interestingly, creation and annihilation do not need to occur between the same pair of critical points. The most important feature is that they must occur between critical points with opposite topological charges, to leave the total topological charge unchanged.

\subsection{Uncharged Case}
Now we turn to study the uncharged case. Setting $q=0$ in Eq. (\ref{eq:tsp_charged_Lovelock}), the spinodal curve reduces to
\begin{eqnarray}
t_{\text{sp}}&=&\frac{d-2}{2 \pi  v \left(6 \alpha  v^2+v^4+15\right)} \Bigl[3 ( d -7 ) + 2 ( d-5 ) \alpha v^{2} \nonumber\\
&&+(d -3)  v^{4}\Bigr]\,, \label{eq:tsp_uncharged_Lovelock}
\end{eqnarray}
and we define
\begin{eqnarray}
f & \equiv& \left(\partial_{v} t_{\text{sp}}\right)_{\alpha} = -\frac{d-2}{2 \pi  v^{2} \left(6 \alpha  v^2+v^4+15\right)^2}\Bigl[45 (d-7)\nonumber\\
&& +4 \alpha  (6 d-57) v^{2}+6 \left(-10 \alpha ^2+2 \alpha ^2 d-5 d+5\right) v^{4}\nonumber\\
&& -12 \alpha  v^{6} +(d-3) v^{8}\Bigr]\,. \label{eq:f_uncharged_Lovelock}
\end{eqnarray}
Analogous to the charged case, this function is continuously differentiable. However, unlike the charged case, we find that this function has different asymptotic behaviors in different dimensions. Specially, when $d=7$, we have
\begin{eqnarray}
f(v \rightarrow 0^+)_{d=7} &\sim& \alpha, \nonumber\\
f(v \rightarrow +\infty)_{d=7}  &\sim& -\frac{ v^{6}}{ \left(6 \alpha  v^2+v^4+15\right)^2} \rightarrow 0^-.
\end{eqnarray} 
While for $d > 7$, it becomes
\begin{eqnarray}
f(v \rightarrow 0^+)_{d > 7} &\sim& -\frac{ (d-2)(d-7)}{ v^2 \left(6 \alpha  v^2+v^4+15\right)^2} \rightarrow -\infty, \nonumber\\
f(v \rightarrow +\infty)_{d > 7}  &\sim& -\frac{ v^{6}}{ \left(6 \alpha  v^2+v^4+15\right)^2} \rightarrow 0^-.
\end{eqnarray}

For positive $\alpha_k$, $\alpha>0$. From Eq. (\ref{eq:one_dimention_degree}), we know that the total topological charge in different dimensions should be given as
\begin{eqnarray}
Q_{\text{total}} &=& \frac{1}{2} [\operatorname{sgn} f(v \rightarrow +\infty)-\operatorname{sgn} f(v \rightarrow 0^+)]\nonumber\\
&=& \left\{\begin{array}{ll}
-1 & \text { for } d = 7 \\
0 & \text { for } d > 7.
\end{array}\right.
\end{eqnarray}
Therefore, the total topological charge of $d=7$ uncharged Lovelock black holes is the same as the ones of charged AdS black holes and charged Lovelock AdS black holes, indicating they can be classified into the same topology class. A detailed analysis shows that the equation of state admits only one critical point and the system demonstrates the small/large black hole phase structure, which may be expected for a same topology class. While the total topological charge of the $d>7$ black holes is different from the $d=7$ ones, indicating they are in different topology classes. Such a difference in topology may imply some distinct differences in their thermodynamics. Indeed, a new phase transition——the reentrant phase transition was found in the $d>7$ case \cite{Xu:2014tja,Frassino:2014pha}. 

\begin{figure}[b]
\centering
\includegraphics[height=5cm]{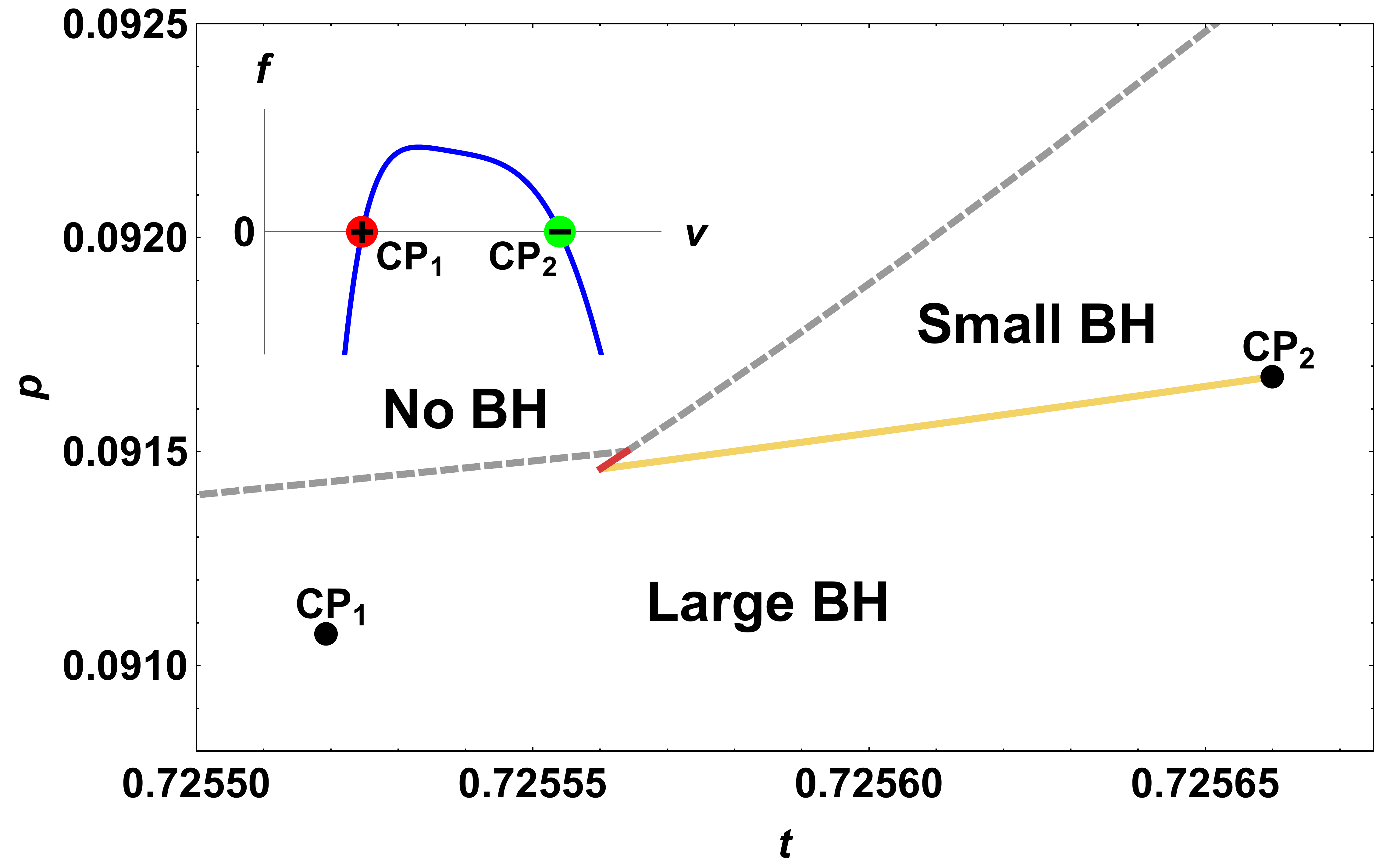}
\caption{\label{fig:P_T_uncharged} Reentrant phase structure for the $d=8$ uncharged Lovelock black hole. Inset: $f \equiv (\partial_{v} t_{\text{sp}})_{\alpha}$ vs $v$ diagram.}
\end{figure}

Taking $d=8$ and $\alpha=2.884$ for example, we show the corresponding phase diagram in Fig. \ref{fig:P_T_uncharged}. Different from the small/large black hole phase transition, the reentrant phase transition consists of a VdW-like first-order phase transition (yellow curve) and a zeroth-order phase transition (red curve). Moreover, two critical points exist in the phase diagram. From the inset displayed in Fig. \ref{fig:P_T_uncharged}, one can see that these critical points possess opposite topological charges, such that
\begin{eqnarray}
Q_{\text{total}}=Q_{\text{CP}_1}+Q_{\text{CP}_2}=0.
\end{eqnarray}

\begin{figure}
\centering
\includegraphics[height=5cm]{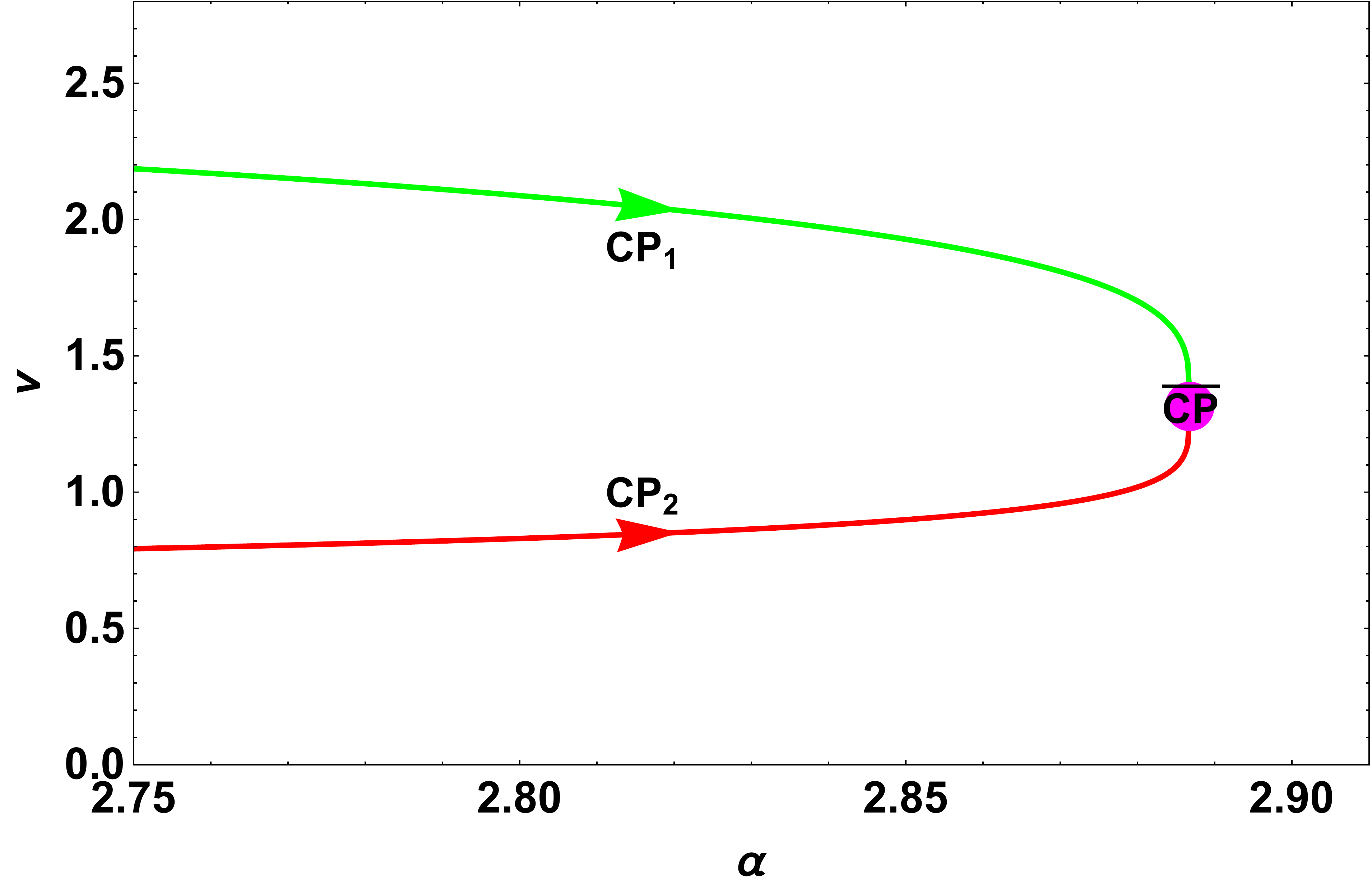}
\caption{\label{fig:Uncharged_Creation} Pair annihilation of critical points for the uncharged AdS Lovelock black hole. The green and red curves represent the branches with negative and positive topological charges, respectively. The arrows refer to the direction of increasing $\alpha$. Two branches intersect at $\alpha_1 \approx 2.886$. We have set $d=8$.}
\end{figure}

On the other hand, due to the invariant total topological charge for $d>7$ black holes, one can expect that there exists pair creation or annihilation of critical points. To observe this behavior, we numerically solve the critical points for $d=8$ black holes with different values of $\alpha$, and then obtain the topological charge for each critical point by reading the corresponding slope of $f$. Results are shown in Fig. \ref{fig:Uncharged_Creation}. It is clear that when $\alpha < \alpha_1 \approx 2.886$, there are two branches of critical points, $\text{CP}_{1}$ and $\text{CP}_{2}$, possessing opposite topological charges. With the increasing of $\alpha$, they get closer to each other. When $\alpha$ goes exactly to $\alpha_1$, two critical points merge into one:
\begin{equation}
\text{CP}_1 + \text{CP}_2 \rightarrow \overline{\mathrm{CP}},    
\end{equation}
where $\overline{\mathrm{CP}}$ has zero topological charge. Further increasing $\alpha$, $\overline{\mathrm{CP}}$ disappears, and there is no critical point anymore.

\section{Topological charge and real critical point}\label{pseudo}

\begin{figure*}
\center{\subfigure[]{\label{P_T_charged_b}
\includegraphics[height=5cm]{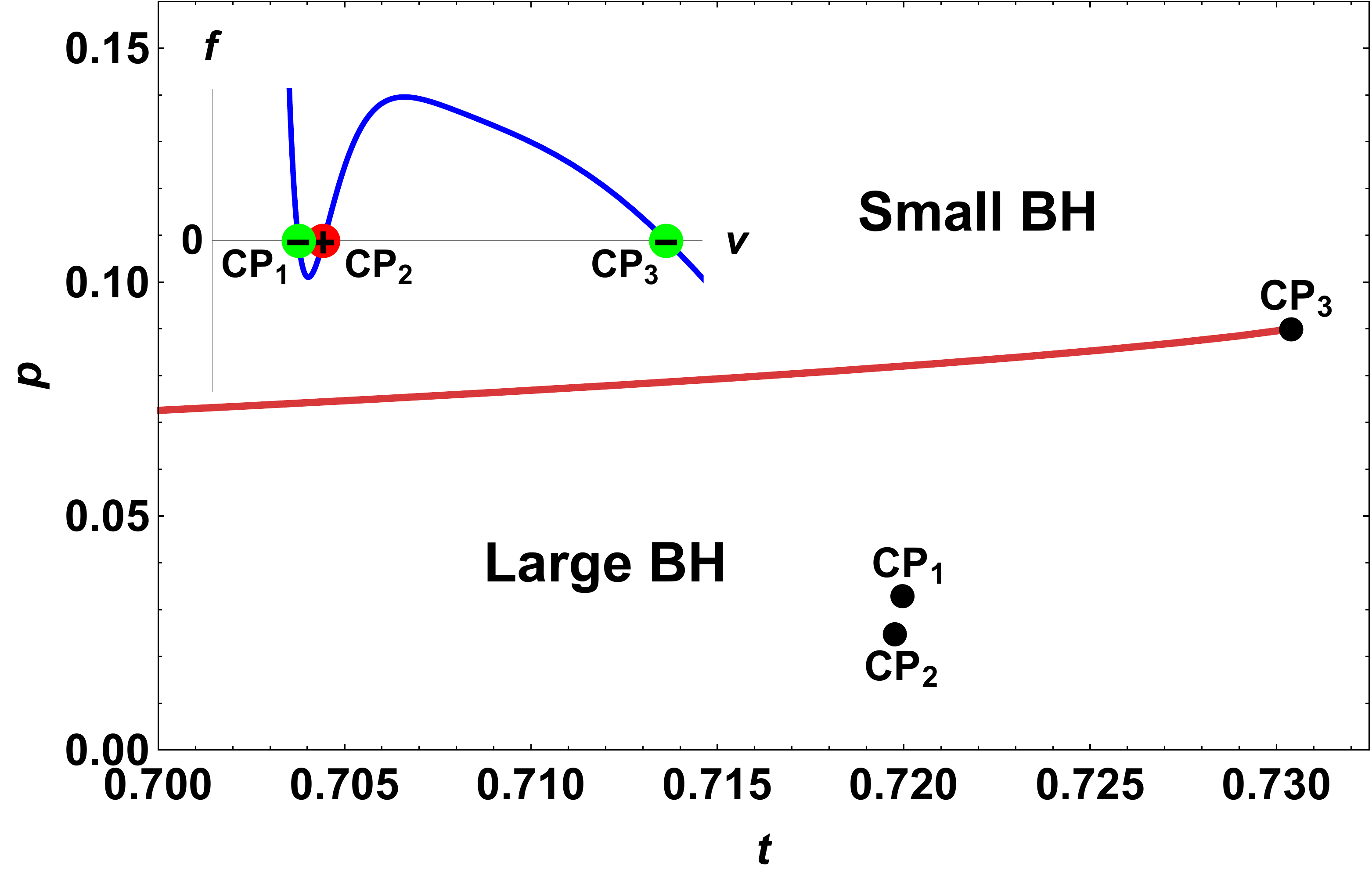}}
\hspace{1em}
\subfigure[]{\label{free_energy}
\includegraphics[height=5cm]{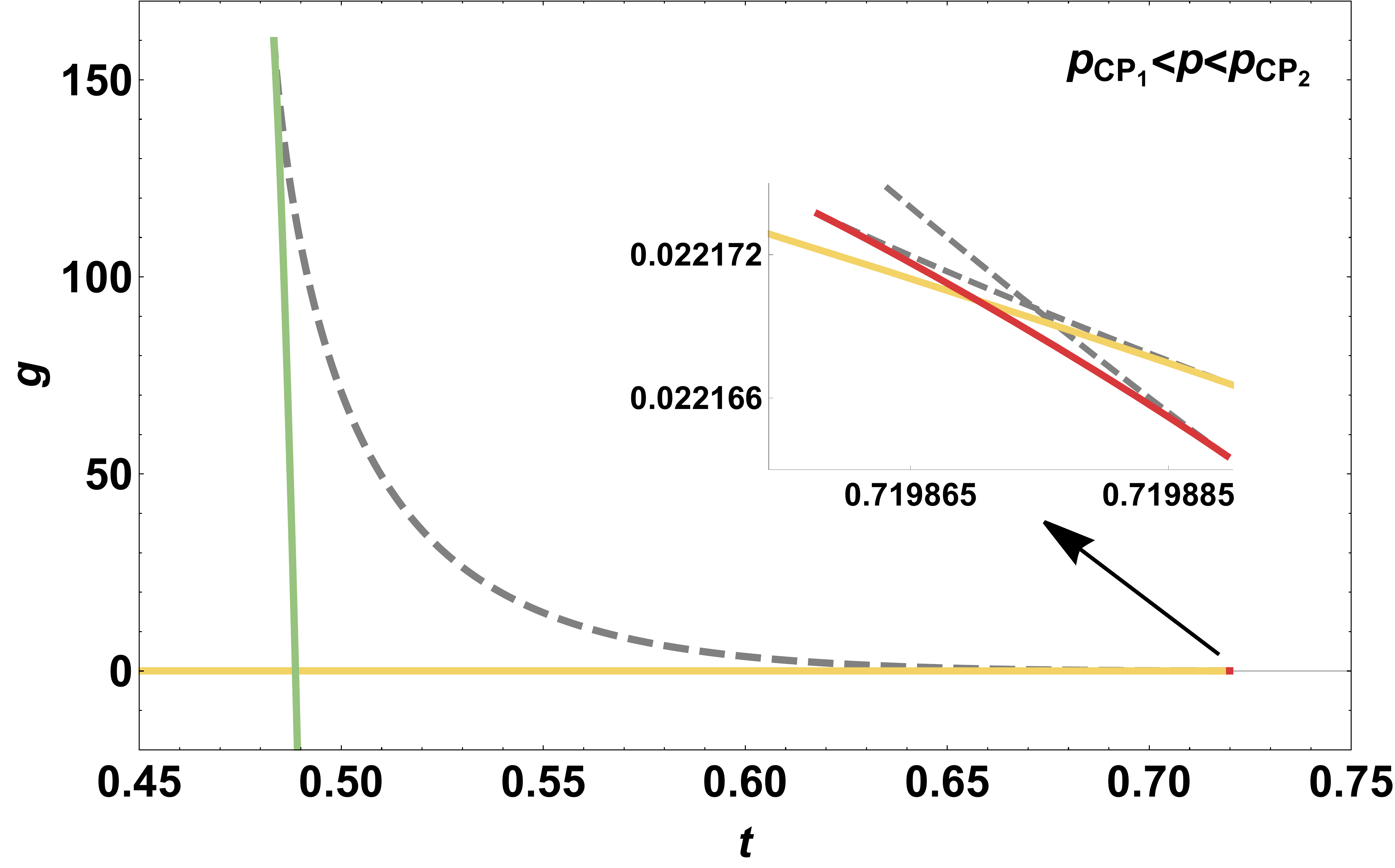}}}
\caption{\label{special_case}(a) Small/large black hole phase structure for the $d=8$ charged Lovelock black hole. (b) $g-t$ diagram. We have set $q=0.021$ and $\alpha=2.8$.}
\end{figure*}

In this section, we would like to discuss the relation between the topological charge and the critical point in more detail. In Sec. \ref{degree}, we have pointed out that a critical point with negative topological charge denotes a maximum point of spinodal curve. Near such a point, using Maxwell's equal law, one can draw a first-order phase transition line in the $T-V$ plane. While for a critical point with positive topological charge, it denotes a minimum point of spinodal curve, and one can not draw a first-order phase transition line near such a point. In Ref. \cite{Wei:2021vdx}, the authors conclude that the first-order phase transition can extend from the conventional critical point (topological charge $-1$), while the presence of the novel critical point (topological charge $+1$) cannot serve as an indicator of the presence of the first-order phase transition near it. In other words, the topological charge can be used to distinguish the 
`\textit{real}' critical point (second-order phase transition point) and the `\textit{pseudo}' one.

For the black holes we studied, this conclusion holds for the vast majority of cases. It is clear that for the phase structures of most interest shown in Figs. \ref{fig:P_T_charged} and \ref{fig:P_T_uncharged}, the conventional critical points indeed connect with the first-order phase transition curve, but the novel ones do not. Nevertheless, we observe a special case in which this conclusion is not applicable, as shown in Fig. \ref{P_T_charged_b}. Such a case occurs in the $d=8$ charged Lovelock AdS black hole, with $q=0.021$ and $\alpha=2.8$. Similar to the small/intermediate/large black hole phase structure, three critical points exist in the system, with topological charges
\begin{eqnarray}
Q_{\text{CP}_1}=-1, \quad Q_{\text{CP}_2}=+1, \quad Q_{\text{CP}_3}=-1.
\end{eqnarray}
However, there are only two stable phases, the large black hole and the small black hole, together with a first-order phase transition curve (red line) between them. We can find that only $\text{CP}_{3}$ connects with the phase transition curve, whereas $\text{CP}_{1}$ and $\text{CP}_{2}$ are below it. Hence, in this case, the topological charge is failing to identify whether a critical point is real or pseudo. In fact, near the critical pressure and temperature of $\text{CP}_{1}$, one can also observe the swallowtail behavior in the Gibbs free energy curve, as shown in the inset of Fig. \ref{free_energy}. But the black hole phases here do not correspond to a minimal energy, and thus $\text{CP}_{1}$ is also a pseudo critical point. It is worth noting that a similar case was also observed in the $d=6$ charged Gauss-Bonnet AdS black holes \cite{Yerra:2022alz}.

Although the negative topological charge may not indicate a real critical point, it still can be a \textit{necessary condition}; the real critical points \textit{can only} emerge from the critical points with negative topological charge.

\section{Conclusions and discussions}\label{conclu}
Starting from the Brouwer degree, we have constructed an approach to probe the topological properties of black hole thermodynamics. The spinodal curve was shown to be a powerful tool to derive the topological information of thermodynamic system. By defining the derivative of spinodal curve as a new function $f$, we can associate a topological charge (degree) for each zero point, i.e., the critical point, and sum over all contributions to construct a total topological charge for the thermodynamic system. Utilizing a mathematical formula, Eq. (\ref{eq:one_dimention_degree}), the analytic calculation of total topological charge is now straightforward; one just needs to examine the asymptotic behavior of spinodal curve's derivative. This enables us to investigate the topological transition between different thermodynamic systems, and to give a topological classification for them conveniently. 

As an example, we first investigated the topology of charged AdS black holes in arbitrary dimensions. We showed that for any $d \geq 4$ and $q >0$, the charged AdS black holes have the same topological charge $-1$, which generalizes the $d=4$, $q=1$ result obtained in Ref. \cite{Wei:2021vdx}. 

Then we turned to study the topology of a more complex system---Lovelock AdS black holes. Specially, we focused on the $\kappa=+1$ case, i.e., the black holes with spherical horizon geometry. For the charged case, it was demonstrated that $d \geq 7$ black holes with arbitrary parameters have the same topological charge $-1$. This indicates that spherical charged Lovelock AdS black holes should be classified into the same topology class with charged AdS black holes, as well as the charged Gauss-Bonnet black holes \cite{Yerra:2022alz}. Thus, it seems that the higher curvature corrections do not change the topology class of black hole thermodynamics in $U(1)$ charged black holes. On the other hand, it is known that these black holes hold two different phase structures---the small/large black hole phase structure and small/intermediate/large black hole phase structure. While from the viewpoint of topology, they are equivalent; the small/intermediate/large black hole phase structure can be interpreted as the topological transformation of the small/large one.

For the uncharged case, we found that the $d=7$ (topological charge $-1$) and $d \geq 8$ (topological charge $0$) black holes are in different topology classes. Such a topological change is found to be accompanied by a change in phase structures---from the small/large black hole phase structure to the reentrant phase structure. Combing the relation between phase structures in the charged case, we conclude that the topological change can be a prognostic indicator of the change in phase structures, but \textit{not vice versa}.

Some general topological properties of critical points were also discussed: i) For a black hole system with non-zero topological charge, there is at least one critical point. ii) For a black hole system with invariant topological charge, critical points must emerge, or annihilate in pairs, between the ones with opposite topological charges. iii) Real critical points can only emerge
from the critical points with negative topological charge. Specially, if a system has total topological charge $-1$, from i), we know that there is at least one critical point. Combing property ii), this explains why there are always odd number of critical points in the system, instead of even number. Analogously, for a system with topological charge $0$, there may be no critical points, or even number of critical points exist. These results confirm the parity conjecture of critical points proposed in Ref. \cite{Yerra:2022alz}, which says that, ``for odd (even) number of critical points, the total topological charge
is an odd (even) number''. Note that the critical points we are talking about here include the ones with negative critical pressure and temperature, and exclude the very special one with topological charge $0$.

It is worth emphasizing that, to obtain the total topological charge, one does not need to get a exact solution for critical points. Some useful information can be directly obtained from this topological quantity, such as the existence and number (odd or even) of critical points, as well as the possible transition in phase structures. This would be quite helpful for the investigation on black hole thermodynamics. 

\begin{acknowledgments}
We are grateful to Yu-Xiao Liu, Jia-Yi Wu, Yu-Chen Huang, Ao-Yun He and Chen Su for useful discussions and valuable comments. This work is supported by NSFC (Grant No.11947408 and 12047573).
\end{acknowledgments}

\end{document}